# Direct Observations of Magnetic Reconnection in the Solar Wind Current Sheets near Mars

Chi Zhang[1,*], Chuanfei Dong[1,2,*], Xinmin Li[1], Han-Wen Shen[3], Jasper Halekas[3], Tai Phan[4], Christian Mazelle[5], Yuki Harada[6], Hongyang Zhou[1], Jiawei Gao[1], Liang Wang[1], Shannon Curry[7], and David L. Mitchell[4]

[1]Center for Space Physics and Department of Astronomy, Boston University, Boston, MA, USA

[2]School of Natural Sciences, Institute for Advanced Study, Princeton, NJ, USA

[3]Department of Physics and Astronomy, University of Iowa, Iowa City, IA, USA

[4]Space Sciences Laboratory, University of California, Berkeley, Berkeley, CA, USA

[5]University of Toulouse, CNES, CNRS, IRAP, Toulouse, France

[6]Department of Geophysics, Graduate School of Science, Kyoto University, Kyoto, Japan

[7]Laboratory for Atmospheric and Space Physics, University of Colorado, Boulder, CO, USA

*Corresponding author: Chi Zhang (zc199508@bu.edu) and Chuanfei Dong (dcfy@bu.edu)

**Abstract**

Magnetic reconnection is a fundamental and ubiquitous process in astrophysical plasmas that converts magnetic energy into plasma kinetic and thermal energy. Throughout the heliosphere, the solar wind is permeated with current sheets (CSs), providing a natural laboratory for investigating this process. Using measurements from NASA's Mars Atmosphere and Volatile EvolutioN (MAVEN) spacecraft, we report the first direct observations of magnetic reconnection occurring within the solar wind CSs near Mars. Specifically, MAVEN observed the classic Petschek-type reconnection exhaust regions, evidenced by bifurcated magnetic field signatures and Alfvénic ion outflows. Notably, the observed exhaust region appears to be large-scale, significantly exceeding the typical thickness of solar wind CSs near Mars. This suggests that magnetic reconnection may significantly broaden the CS. Our results underscore the ubiquity of magnetic reconnection across heliocentric distances and may provide new insights into the large-scale evolution of the solar wind and the development of turbulence within it.

## 1 Introduction

Magnetic reconnection is a fundamental plasma process that occurs in CSs, where it reconfigures magnetic field topology and converts stored magnetic energy into plasma kinetic and thermal energy (e.g., Ji et al., 2022; Paschmann et al., 2013; Hesse and Cassak, 2020). It has been widely observed across diverse space and astrophysical environments, including the solar atmosphere, solar wind, Earth's magnetosphere, and other planetary magnetospheres, and plays a key role in particle acceleration as well as mass and energy transport (e.g., Angelopoulos et al., 2008; Burch & Nakamura, 2025; Drake et al., 2025; Nakamura et al., 2025; Gershman et al., 2024; Dong et al., 2018, 2022; Øieroset et al., 2001; Harada et al., 2015; Li et al., 2022; T. L. Zhang et al., 2012).

Within the heliosphere, the solar wind provides an ideal natural laboratory for studying magnetic reconnection because it is permeated by numerous CSs (e.g., Gosling, 2011). Generally, solar wind CSs can generally be classified into two categories. The first type is the large-scale heliospheric current sheet (HCS), which arises from polarity reversals of the interplanetary magnetic field (IMF) (Smith et al., 1978). The second type consists of smaller-scale, transient, and randomly distributed CSs (often referred to as solar wind discontinuities), which may either originate near the Sun and be transported outward or form through magnetohydrodynamic (MHD) turbulence (e.g., Greco et al., 2009). Observationally, these two types of CS can be distinguished by their magnetic field and electron signatures. The HCS is typically identified by a reversal in the radial component of the IMF, often accompanied by a reversal in the strahl electron (typically in the 100 eV to 1 keV range) pitch angle distribution from predominantly parallel to predominantly antiparallel, or vice versa (e.g., Szabo et al., 2020). In contrast, other CSs within the solar wind are categorized as random CSs, which generally exhibit more heterogeneous magnetic field and electron signatures (e.g., Phan et al., 2021).

Direct evidence of magnetic reconnection within the solar wind CS has been extensively reported across various heliocentric distances. The first in-situ spacecraft observations of solar wind magnetic reconnection were reported by Gosling et al. (2005a) at Earth's orbit (1 AU, ~$1.5 \times 10^8$ km). And it is found that the reconnection can occur in both HCS (e.g., Ala-Lahti et al., 2025; Gosling et al., 2005b) and random CS (Phan et al., 2006, 2009; Eriksson et al., 2022; Øieroset et al., 2017). More recently, with the advent of the Parker Solar Probe, magnetic reconnection has also been identified in the near-Sun environment (e.g., Phan et al., 2020, 2024), indicating that reconnection is highly prevalent throughout the inner heliosphere (Phan et al., 2009, 2021). Furthermore, magnetic reconnection is not limited to the inner heliosphere; based on Ulysses observations, Gosling et al. (2006) demonstrated that reconnection can still actively occur in the outer heliosphere (1.4–5.4 AU).

It is worth noting that many reported reconnection events are identified when spacecraft traverse quasi-stationary, large-scale exhaust regions. These exhausts are typically characterized by Alfvénic ion outflows bounded by bifurcated magnetic field signatures (Gosling and Szabo, 2008). Specifically, such signatures consist of two distinct magnetic field rotations at the edges of the CS, separated by a relatively flat region in between. This double-step structure indicates that the original CS has bifurcated into two boundaries enclosing a central exhaust region, consistent with the Petschek reconnection model, in which the exhaust is bounded by a pair of Alfvén or slow-mode waves (Petschek, 1964). Observations of reconnection exhaust regions are relatively common owing to their large spatial scales, which can extend over thousands of ion inertial lengths ($d_i = c/\omega_{pi}$), where $c$ is the speed of light and $\omega_{pi}$ is the ion plasma frequency. In contrast, direct observations of the diffusion region remain exceedingly rare due to its much smaller scale ($\sim$ 1–10 $d_i$). Only recently, enabled by high-resolution spacecraft measurements, have the ion and electron diffusion regions been directly resolved in the solar wind (Wang et al., 2022, 2023).

While solar wind reconnection has been widely detected across the heliosphere, direct observations of this phenomenon near Mars have not yet been reported. Unlike Earth, Mars lacks a global intrinsic magnetic field and hosts an induced magnetosphere instead (e.g., Cheng et al., 2026; Zhang et al., 2022, 2025a, 2026), representing a unique plasma environment between the pristine solar wind and magnetized planetary systems. Previous studies have shown that solar wind CSs are frequently encountered at Mars (Chen et al., 2024) and can significantly alter the global structure of the induced magnetosphere (Wen et al., 2025; Guo et al., 2026). However, these observed CSs have been primarily identified as non-reconnecting. Consequently, it remains unknown whether and how magnetic reconnection develops and manifests within the solar wind under these specific Martian conditions.

In this study, we utilize MAVEN observations to present direct evidence of magnetic reconnection within the solar wind random CSs at Mars' orbit. Our findings suggest that magnetic reconnection is a ubiquitous and robust process throughout the heliosphere. These results may offer new insights into the large-scale evolution of the solar wind and the fundamental role of reconnection in governing heliospheric dynamics.

## 2. Data Sets and Methodology

We use magnetic field measurements from the Magnetometer (MAG; Connerney et al., 2015), ion data from the Solar Wind Ion Analyzer (SWIA; Halekas et al., 2015), and electron measurements from the Solar Wind Electron Analyzer (SWEA; Mitchell et al., 2016), all onboard MAVEN (Jakosky et al., 2015). MAG is a fluxgate magnetometer that measures three-dimensional magnetic field vectors at sampling frequencies of 32 Hz and 1 Hz. In this study, we use the 1 Hz magnetic field data. SWIA measures ions over an energy range from 25 eV/q to 25 keV/q. The full three-dimensional ion distributions provided by SWIA have a cadence of 8 s and a field of view of $360° \times 90°$. SWEA measured full three-dimensional electron distributions over an energy range from 3 eV/q to 46 keV/q with a time resolution of 2 s. SWEA has a field of view of $360° \times 120°$, with approximately 8% of the viewing area obscured by the spacecraft.

All measurements are analyzed in the Mars Solar Orbital (MSO) coordinate system, where the $\vec{X}_{MSO}$ is along the vector from Mars to the Sun, $\vec{Z}_{MSO}$ is perpendicular to the orbital plane, and $\vec{Y}_{MSO}$ completes the right-handed system, closely aligned with the opposite direction of the orbital velocity vector.

To investigate the properties of the CSs, we further construct a local LMN coordinate system using a hybrid minimum variance method (e.g., Gosling & Phan, 2013; Phan et al., 2020). The CSs normal direction, $\vec{N}_{CS}$, is determined from $\vec{B}_1 \times \vec{B}_2 / | \vec{B}_1 \times \vec{B}_2 |$, where $\vec{B}_1$ and $\vec{B}_2$ are the magnetic fields on the two sides of the CSs. The $\vec{M}_{CS}$ is then defined as, $\vec{M}_{CS} = \vec{N}_{CS} \times \vec{L}_{MVA}$, which is approximately aligned with the out-of-plane X-line direction. Here, $\vec{L}_{MVA}$ is the maximum variance direction obtained from minimum variance analysis (MVA) of the magnetic field (MVAB; Sonnerup & Scheible, 1998). Finally, the third axis is given by $\vec{L}_{CS} = \vec{M}_{CS} \times \vec{N}_{CS}$, thereby completing a right-handed LMN system and approximately corresponding to the reconnecting (antiparallel) magnetic field direction.

## 3. Observations

### 3.1 Event 1: 15 June 2025

Between 12:00 and 12:10 UTC on 15 June 2025, MAVEN was located at approximately (2.21, 0.57, -0.33) $R_M$ (Figures 1a–1b), well upstream of the nominal bow shock. The ion energy spectra measured by SWIA exhibit two narrow, beam-like populations centered at ~1.5 keV and ~3 keV (Figure 1c), corresponding to $H^+$ and $He^{++}$, respectively. The presence of these well-defined beams indicates a cold solar wind, confirming that MAVEN was indeed in the upstream region.

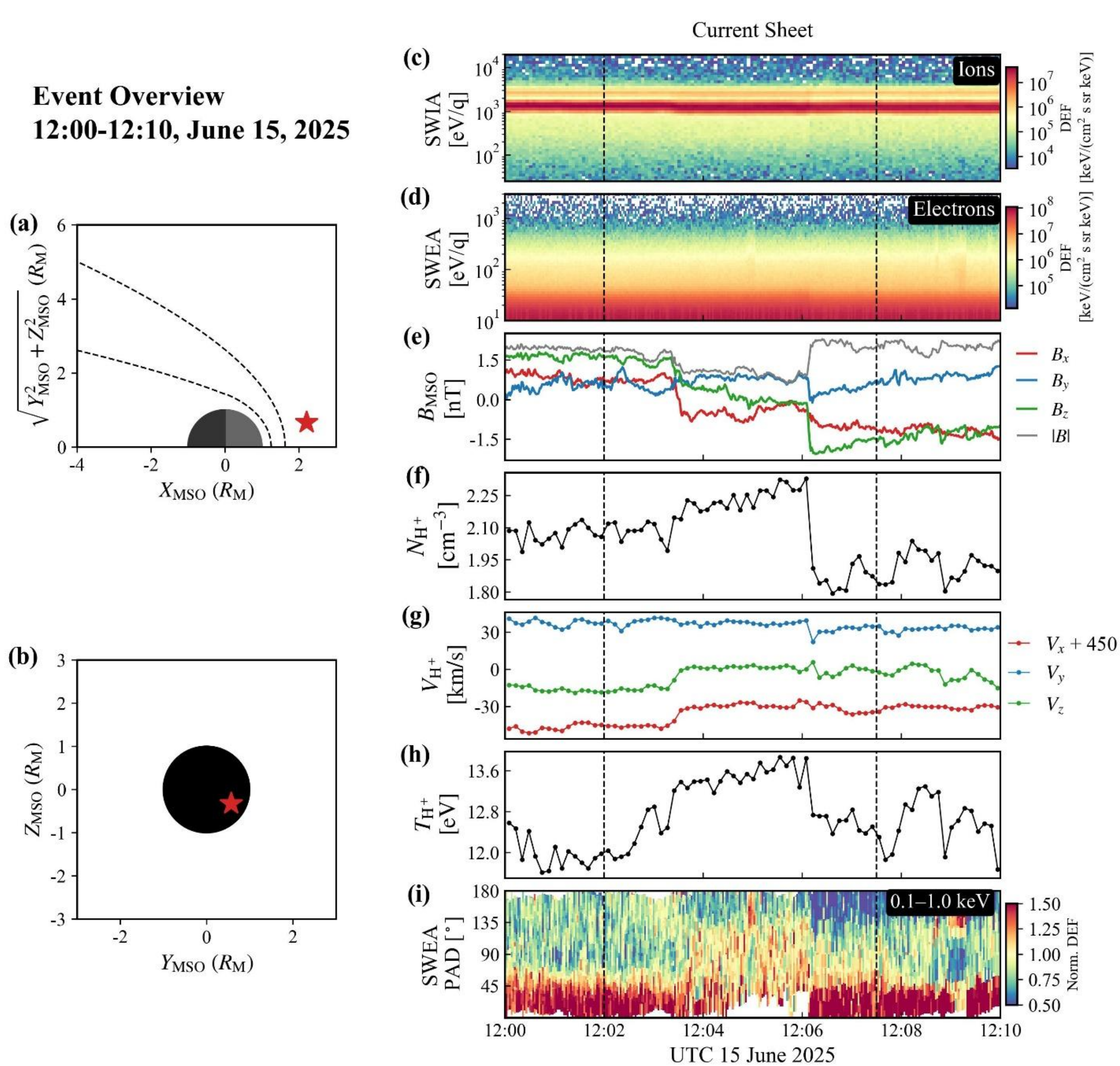

**Figure 1**. **Overview of the event during 15, June 2025.** (a) MAVEN location in the $X_{\mathrm{MSO}} - \sqrt{Y_{\mathrm{MSO}}^2 + Z_{\mathrm{MSO}}^2}$ plane. (b) MAVEN location in the $YZ_{\mathrm{MSO}}$ plane. (c) Ion energy spectrogram measured by SWIA. (d) Electron energy spectrogram measured by SWEA. (e) Magnetic field components in MSO coordinates. (f) Proton number density. (g) Proton bulk velocity in MSO coordinates. (h) Proton temperature. (i) Electron pitch-angle distribution for energies between 100 eV and 1 keV.

During 12:02–12:07:30 UT (marked by the black vertical dashed lines in Figures 1c–1i), the IMF underwent a clear rotation (Figure 1e). Specifically, the magnetic field changed from approximately $\vec{B}_1 = (0.85, 0.53, 1.64)$ nT during 12:00-12:02 UT to $\vec{B}_2 = (-1.21, 0.86, -1.31)$ nT during 12:07:30–12:10 UT. At the center of the transition (12:04-12:06), the magnetic field magnitude ($|B|$) decreased to ~1 nT. These signatures indicate a CS crossing, with a magnetic shear angle of ~135.5°.

During the crossing, the solar wind proton bulk parameters also exhibit systematic variations. The proton density is ~2.05 $cm^{-3}$ on the left side of the CS, increases to ~2.3 $\mathrm{cm}^{-3}$ near the center, and then decreases to ~1.8 $\mathrm{cm}^{-3}$ on the right side (Figure 1f). Accordingly, the $d_i$ ranges from ~150 to ~170 km. Meanwhile, the proton bulk velocity also shows a slight reduction, particularly in the negative $V_x$ component, decreasing from ~−500 km/s on the left side to ~−470 km/s near the center (Figure 1g). In addition, the proton temperature increases from ~12 eV on the left side to ~14 eV near the center, before decreasing slightly to ~12.5 eV on the right side (Figure 1h). These results indicate that the CS separates two asymmetric plasma regions.

This CS is unlikely to be HCS, which typically exhibits a clear $B_x$ reversal and a corresponding reversal in the pitch-angle distribution of strahl electron (typically in the 100 eV to 1 keV range). In general, the strahl electrons are field-aligned on one side of the HCS and anti-field-aligned on the other (i.e., parallel/antiparallel or vice versa). However, no such reversal in the strahl electron population is observed in this event (Figure 1i); instead, the strahl electrons remain primarily parallel to the magnetic field on both sides of the crossing. In addition, the dominant magnetic field rotation occurs in the $B_z$ component rather than $B_x$. These characteristics suggest that this structure is more likely a local, or "random" CS.

The local LMN coordinate system is constructed using the magnetic field vectors on the two sides of the CS ($\vec{B}_1$ and $\vec{B}_2$), combined with the MVA applied to the magnetic field over the interval 12:02–12:07:30 (see Figure 2a). The MVA results give $\vec{L}_{MVA} = (0.46, 0.05, 0.89)$, with eigenvalue ratios of $\lambda_1 : \lambda_2 : \lambda_3 = 1.77 : 0.12 : 0.02$. The ratio $\lambda_1/\lambda_2 = 14.5$ indicates that $\vec{L}_{\mathrm{MVA}}$ is well determined. The CS normal direction is determined from the $\vec{B}_1 \times \vec{B}_2 / |\vec{B}_1 \times \vec{B}_2|$, yielding $\vec{N}_{CS} = (-0.79, -0.32, 0.52)$. Using $\vec{N}_{\mathrm{CS}}$ and $\vec{L}_{\mathrm{MVA}}$, we obtain $\vec{M}_{\mathrm{CS}} = \vec{N}_{\mathrm{CS}} \times \vec{L}_{\mathrm{MVA}} = (-0.31, 0.94, 0.11)$, and thus $\vec{L}_{\mathrm{CS}} = \vec{M}_{\mathrm{CS}} \times \vec{N}_{\mathrm{CS}} = (0.52, 0.08, 0.85)$.

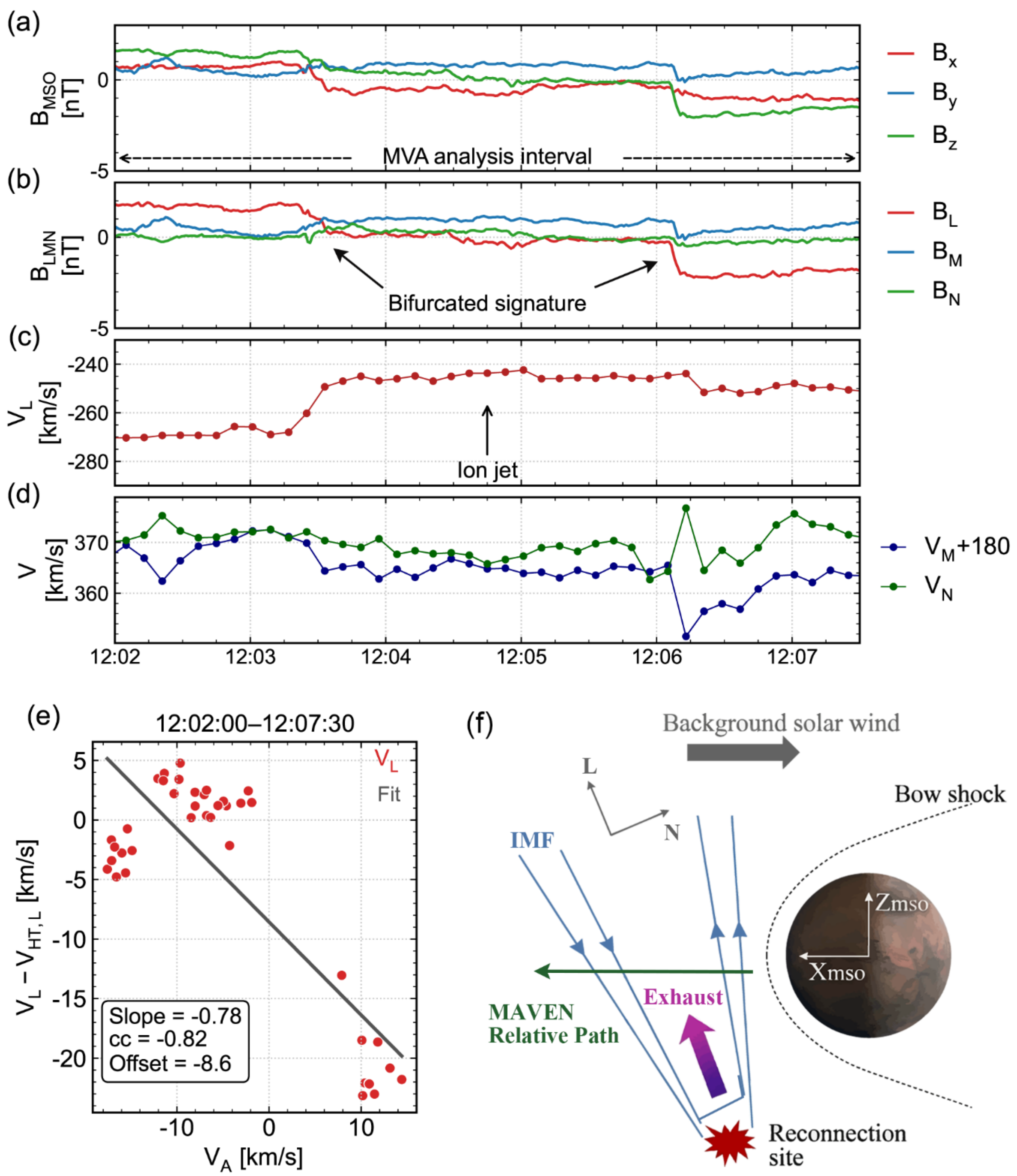

**Figure 2.** (a) Magnetic field components in MSO coordinates. (b) Magnetic field components in LMN coordinates, showing a bifurcated signature in $B_L$. (c) Ion bulk velocity $V_L$, exhibiting a clear ion jet. (d) Ion velocity components $V_M$ and $V_N$. (e) Walén test comparing plasma velocity and Alfvén velocity, with the linear fit shown. (f) Schematic illustration of the event, depicting the spacecraft trajectory relative to the background solar wind flow, together with the reconnection exhaust, and reconnection site (not to scale).

The observed magnetic field and ion velocity are then rotated into the LMN coordinates. In the LMN frame, Figure 2b shows that the $B_L$ component exhibits a clear bifurcated structure: it initially decreases from ~2 nT to ~0 nT, then remains near zero in the central region, and finally decreases to ~ −2 nT. Concurrently, the ion bulk velocity $V_L$ decrease from ~ −270 km/s on the left side to ~ -242 km/s near the center of the CS (Figure 2c), and further changes to ~ -250 km/s as the spacecraft exits the CS on the right side.

The hybrid Alfvén speed ($V_A$) was estimated based on the reconnecting magnetic field component $B_L$ and the proton density $n$ on the two sides of the CS, following Cassak & Shay (2007):

$$V_A = \sqrt{\frac{B_{L1}B_{L2}(B_{L1}+B_{L2})}{\mu_0 m_p(n_1 B_{L2}+n_2 B_{L1})}} \quad (1)$$

where $B_{L1}$ and $B_{L2}$ are the reconnecting magnetic field components on the two sides of the CS, $n_1$ and $n_2$ are the corresponding proton number densities, $m_p$ is the proton mass, and $\mu_0$ is the vacuum permeability. Using $B_{L1} = B_{L2} = 2$ nT, $n_1 = 2.05\ cm^{-3}$, and $n_2 = 1.85\ cm^{-3}$, the hybrid Alfvén speed ($V_A$) is estimated to be ~ 31 km/s. Thus, the observed variations in $V_L$ are ~0.9 $V_A$ on the left side and ~0.26 $V_A$ on the right side, consistent with Alfvénic reconnection outflows. To further assess this interpretation, we perform a Walén test over the interval from 12:02 to 12:07:30. The test yields a correlation coefficient of $\sim -0.82$ (Figure 2d), indicating that the plasma outflows are predominantly Alfvénic on average.

We further analyze the reduced ion velocity distributions shown in Figure A1 of Appendix A. As illustrated in Figures A1a–A1d, both the one-dimensional and two-dimensional reduced distributions show that the proton $V_L$ within the exhaust region shifts systematically toward smaller values, confirming the presence of the Alfvénic outflow. This also suggests that reconnection drives the bulk deceleration along $\vec{L}_{CS}$.

In addition, the ion temperature in the exhaust region is enhanced by ~1.6 eV and ~1.0 eV relative to the left and right sides, respectively (Figure 1i). This enhancement is consistent with ion bulk heating associated with magnetic reconnection. Empirically, reconnection-driven ion heating is often found to scale as $0.13\, mV_A^2$, which yields an expected temperature increase of ~1.3 eV (Phan et al., 2014; Øieroset et al., 2024), in good agreement with the observations.

We also note that the electron pitch-angle distribution within the exhaust region differs from those on the left and right sides (Figure 1i). To further investigate this feature, we analyze the electron energy–pitch angle distributions shown in Figure B1 of Appendix B. On both the left and right sides of the exhaust region, electrons in the energy range of 30–300 eV are predominantly field-aligned, indicating open magnetic field lines connected to the Sun. In contrast, the electron distribution within the exhaust region becomes much more isotropic, indicating possible modification of the magnetic topology.

Together with the bifurcated magnetic field structure, Alfvénic exhaust, and bulk ion heating, these results provide strong evidence that MAVEN observed a Petschek-type reconnection exhaust. Furthermore, the average $|B_M|$ is ~0.74 nT, implying a guide field ratio $|B_M/B_L| \approx 0.37$, indicative of moderate guide-field reconnection.

Based on the local coordinate system of the CS, together with the magnetic field and ion velocity signatures, we propose the schematic interpretation shown in Figure 2f. In the LN plane, the background solar wind flows predominantly in the $-\vec{L}$ and $+\vec{N}$ directions, whereas the reconnection exhaust is directed mainly along $+\vec{L}$. Although the exhaust flow is Alfvénic, its speed is smaller than that of the ambient solar wind. Since the spacecraft velocity is negligible compared with the solar wind speed, MAVEN effectively samples the CS as it convects past the spacecraft with the background flow. Under this geometry, the exhaust flow partially offsets the background solar wind velocity in the $-\vec{L}$ direction, leading to the observed reduction of the ion flow along $-\vec{L}$.

### 3.2 Event 2: 12 January 2015

Figure 3 shows the second representative reconnection CS observed in the solar wind between 16:40 and 16:50 UT on 12 January 2015, when MAVEN was located at approximately (1.88, 0.14,

−2.09) $R_M$ (Figures 3a–3b). The SWIA measurements reveal a narrow, beam-like ion population centered at ~600 eV (Figure 3c), corresponding to $H^+$. A weaker $He^{2+}$ population is also present, although less clearly defined, at around ~1 keV. These signatures further confirm that MAVEN was in the upstream solar wind.

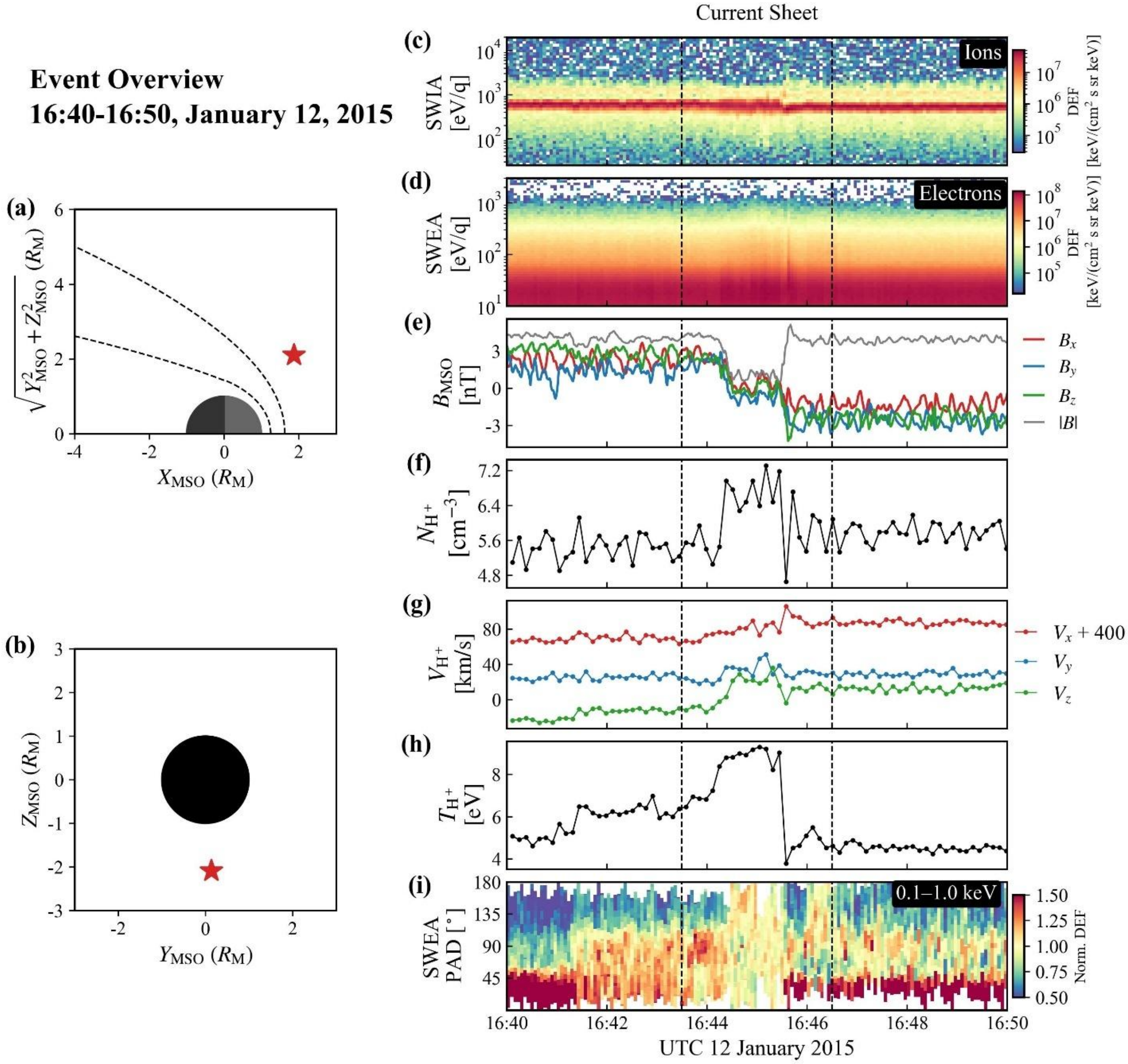

**Figure 3**. **Overview of the event during 12 January 2015.** The layout is consistent with that of Figure 1.

Similar to the first case, MAVEN detected a clear IMF rotation during 16:43:30–16:46:30 UT (Figure 3e), indicating a current sheet crossing (marked by the black vertical dashed lines in

Figures 3c–3i). The magnetic field vector rotated from $\vec{B}_1 = (2.24, 1.35, 2.93)$ nT during 16:40-16:43 UT to $\vec{B}_2 = (-1.37, -2.58, -2.43)$ nT during 16:47-16:50 UT, corresponding to a shear angle of ~156.4°. At the center of the transition (~16:45), the magnetic field magnitude |B| decreased to ~1.3 nT. Unlike the first case, this interval exhibits quasi-periodic magnetic field fluctuations throughout the entire time interval, with a characteristic period of ~20 s. This timescale is comparable to the local proton gyroperiod, suggesting the presence of proton cyclotron waves (e.g., Zhang et al., 2025b). These waves are generated by newly ionized planetary protons originating from the extended hydrogen exosphere beyond the bow shock (e.g., Chaffin et al., 2015).

The plasma parameters across the CS exhibit slight asymmetries. The proton density increases from ~5.4 $cm^{-3}$ on the left side to ~7.2 $cm^{-3}$ near the center, and then decreases to ~5.7 $cm^{-3}$ on the right side (Figure 3f), corresponding to a $d_i$ of ~85–100 km. Meanwhile, the proton temperature increases from ~6 eV on the left side to ~9 eV near the center, before decreasing to ~4.8 eV on the right side (Figure 3h). Similar to the first case, Figure 3i shows that, during the crossing, the strahl electrons remain predominantly parallel to the magnetic field on both sides, indicating that this CS also belongs to a random CS rather than HCS.

Applying the same method as for the first case, we construct the local LMN coordinate system using the magnetic field vectors on the two sides of the CS, combined with the MVA applied to the magnetic field over the interval 16:43:30–16:46:30 (Figure 4a). The resulting CS normal direction is $\vec{N}_{CS} = (0.72, 0.24, -0.66)$. The MVA yields $\vec{L}_{MVA} = (0.54, 0.6, 0.62)$, with eigenvalue ratios of $\lambda_1 : \lambda_2 : \lambda_3 = 9.24 : 0.33 : 0.23$. The large ratio $\lambda_1/\lambda_2 = 27.8$ also indicates that $\vec{L}_{\mathrm{MVA}}$ is well determined. Using $\vec{N}_{\mathrm{CS}}$ and $\vec{L}_{\mathrm{MVA}}$, we derive $\vec{M}_{CS} = (0.54, -0.79, 0.30)$, and $\vec{L}_{\mathrm{CS}} = (0.45, 0.56, 0.69)$.

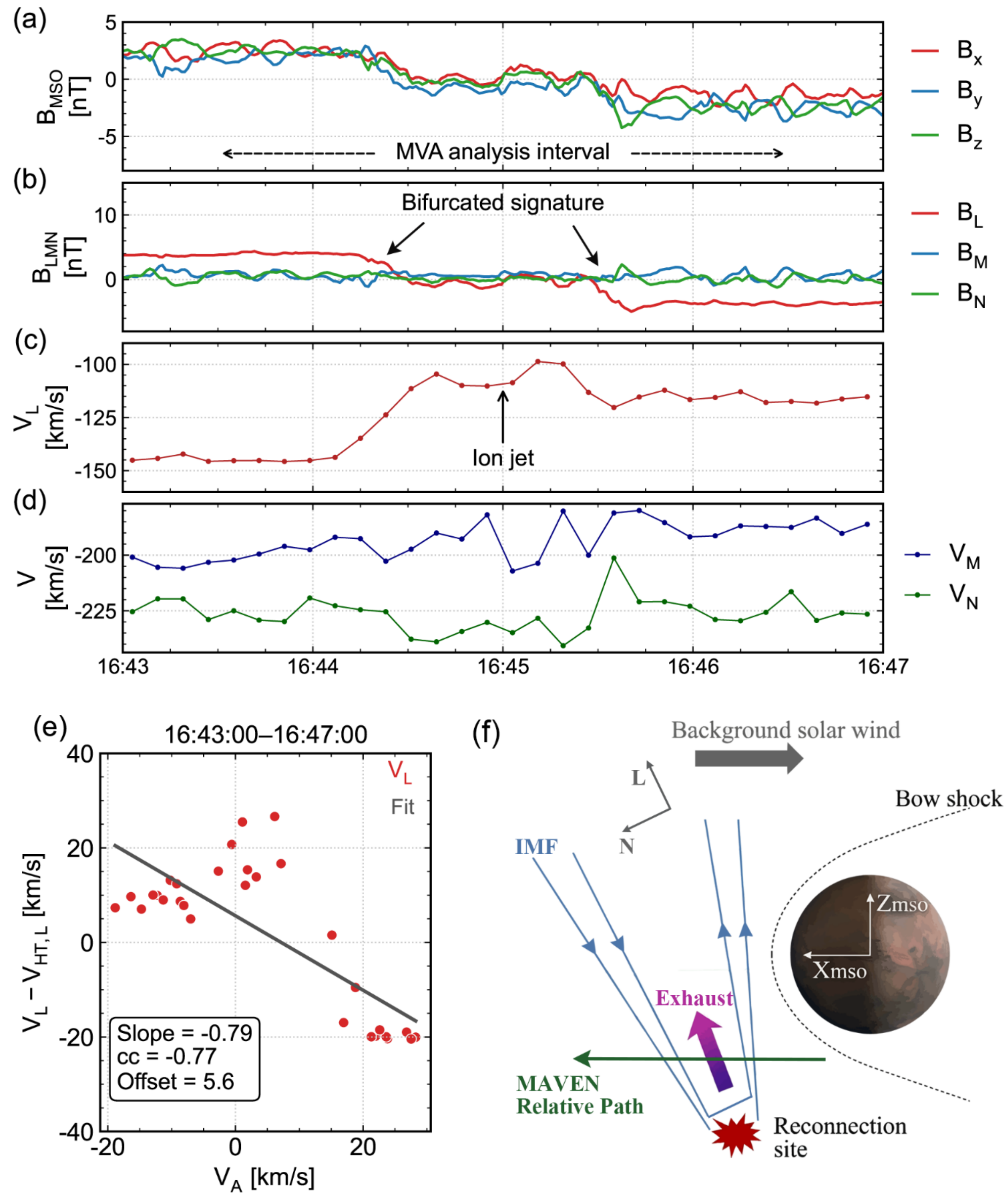

**Figure 4.** The format is the same as in Figure 2, but applied to the event on 12 January 2015.

The $B_L$ exhibits a clear bifurcated structure across the CS (Figure 4b). Concurrently, the ion velocity component $V_L$ shows a localized enhancement within the CS (Figure 4c), indicative of an ion jet. Specifically, $V_L$ increases from $\sim -144$ km/s on the left side to $\sim -110$ km/s near the center, and then decreases to $\sim -120$ km/s on the right side. Using the parameters of $B_{L1} = 3.93$ nT and $B_{L2} = 3.8$ nT, and proton densities of $n_1 = 5.4\ cm^{-3}$ and $n_2 = 5.7\ cm^{-3}$, the corresponding hybrid Alfvén speed is $V_A = 35.77$ km/s. Thus, the velocity variations correspond to $\sim 0.96\,V_A$ on the left side and $\sim 0.28\,V_A$ on the right side, consistent with Alfvénic outflows. The Walén test yields a correlation coefficient of ~ -0.77, further supporting the Alfvénic nature of the jet (Figure 4e). Similar to the first case, the ion velocity distributions shown in Figures A1e–A1h also suggest that the observed variation in $V_L$ within the exhaust region is caused by reconnection-driven bulk deceleration.

Based on empirical scaling, reconnection-driven ion heating is expected to produce a temperature increase of ~1.74 eV. The observed temperature enhancements are ~3 eV and ~4.2 eV on the two sides of the current sheet, respectively. Although these values are noticeably higher than the predicted value, they remain within the scatter of the dataset used to derive the empirical relation (see Figure 2g in Phan et al., 2014). This suggests that the observed heating is still broadly consistent with reconnection-driven processes.

These signatures also strongly support the interpretation that the spacecraft observed a Petschek-type reconnection exhaust, as illustrated in Figure 4f. In the LN plane, the background solar wind flows predominantly in the $-\vec{L}$ and $-\vec{N}$ directions, whereas the reconnection exhaust is directed along $+\vec{L}$. Thus, the exhaust flow partially cancels the background solar wind velocity in the $-\vec{L}$ direction, resulting in the observed reduction in the ion flow along $-\vec{L}$.

Similar to the first case, the electron pitch-angle distribution within the exhaust region becomes more isotropic (Figure B1), whereas the distributions on the left and right sides remain predominantly field-aligned. This result also suggests that the left and right sides correspond to open magnetic field lines, while reconnection within the exhaust region may modify the magnetic topology.

## 4. Discussion

The variation in the background solar wind velocity induced by reconnection outflows observed at Mars is relatively small compared to those reported near the Sun or at 1 AU. This difference primarily reflects the radial evolution of solar wind parameters. The local Alfvén speed at Mars' orbit is about 30 km/s, lower than the ~50 km/s at 1 AU and the several hundred km/s near the Sun (e.g., Phan et al., 2022). Accordingly, the observed outflow speeds of ~10–30 km/s are consistent with the expected Alfvénic scaling, indicating that the fundamental physics of magnetic reconnection remains similar across different heliocentric distances, despite large variations in the absolute plasma parameters.

The CS thickness $(D)$ can be roughly estimated as $D = | V_N | \cdot \delta t$, where $\delta t$ is the crossing duration. Using this method, the thickness of the first CS is estimated to be $\sim 58400$km ($\sim 17.2\, R_M$) or $\sim 365\, d_i$, while the second case has an estimated thickness of $\sim 22500$km, corresponding to $\sim 6.6\, R_M$ or $\sim 250\, d_i$. These values are significantly larger than the typical thickness of solar wind CS observed near Mars, which is usually $\sim 30\, d_i$ or $\sim 1–2\, R_M$ (Chen et al., 2024). Such large scales are physically plausible because magnetic reconnection can progressively broaden the exhaust region as it propagates away from the reconnection site (Phan et al., 2009). For the first event, additional evidence for its large spatial extent is provided by observations from the Tianwen-1 Mars Orbiter Magnetometer (Wan et al., 2020; Y. Wang et al., 2023; G. Wang et al., 2024; Zou et al., 2023). During the interval, Tianwen-1 was located in the nightside flank magnetosheath, approximately $4.5\, R_M$ away from MAVEN (Figure C1 in the Appendix C). Despite this large separation, Tianwen-1 observed magnetic field signatures similar to those detected by MAVEN between 12:04 and 12:08 UT, including a reversal of the $B_z$ component (Figure C1d) and a similar evolution of the clock angle (Figure C1e). The detection of same CS at two locations separated by several Martian radii indicates that the CS extended over at least this scale, providing direct support for its exceptionally large spatial extent. Notably, the inferred spatial scale of the reconnection exhausts also significantly exceeds the characteristic size of the Martian induced magnetosphere, which is only a few $R_M$ (e.g., Zhang et al., 2022). This suggests that magnetic reconnection within solar wind CSs may influence the Martian plasma environment on a global scale.

Previous studies have shown that solar wind CS, through their associated IMF rotations, can significantly alter the global configuration of the Martian induced magnetosphere (e.g., Wen et al., 2025; Guo et al., 2026). In addition to this effect, magnetic reconnection within the CS may further modulate the solar wind flow through local ion acceleration or deceleration. Although the reconnection-driven velocity perturbations are small compared with the background solar wind speed, they should still produce corresponding variations in the solar wind dynamic pressure. Variations in the dynamic pressure are known to strongly affect the Martian plasma environment, including the large-scale magnetic field configuration (e.g., Zhang et al., 2022), ion dynamics (e.g., Dubinin et al., 2017), and electron distributions (e.g., Zhang et al., 2025c). To estimate the magnitude of this effect, we calculated the dynamic pressure on both sides of the exhaust region and within the exhaust itself. For the 2025 June 15 event, the average dynamic pressure is 0.868 nPa on the left side, 0.878 nPa in the exhaust region, and 0.736 nPa on the right side. Thus, the exhaust value is enhanced by about 9.5% relative to the average of the two adjacent sides. For the 2015 January 12 event, the corresponding values are 1.003 nPa, 1.153 nPa, and 0.944 nPa, respectively, giving an enhancement of about 18.4% within the exhaust region. These estimates indicate that the reconnection-caused dynamic pressure variations are relatively modest.

In addition to dynamic pressure variations, reconnection-driven plasma heating may also modify the thermal pressure of the solar wind. However, the observed ion temperature within the exhaust region is not significantly enhanced, indicating that the reconnection-associated heating in both events is relatively weak. This result is also consistent with established empirical reconnection scaling relations for ion bulk heating, suggesting that the heating is limited by the relatively low local Alfvén speed. Therefore, magnetic reconnection produces only relatively modest perturbations to the solar wind ion properties in both events.

## 5. Conclusion

In summary, we report the first observations of magnetic reconnection in solar wind current sheets near Mars. Specifically, MAVEN directly sampled Petschek-type reconnection exhausts, characterized by Alfvénic ion jets within bifurcated CSs and ion bulk heating consistent with established reconnection scaling relations. The spatial scales of the reconnection exhausts are significantly larger than those of typical solar wind CSs near Mars. This is consistent with expectations, as reconnection bifurcates the original CS and leads to continuous broadening of the exhaust as it convects away from the reconnection site. Analysis of the electron pitch-angle distributions further indicates that the two analyzed CSs are not associated with the HCS, but instead correspond to random CSs embedded within the solar wind, likely generated by MHD turbulence. In addition, the electron pitch-angle distributions become more isotropic within the exhaust regions, while remaining predominantly field-aligned in the surrounding regions, suggesting that reconnection may modify the magnetic topology.

Our results demonstrate that magnetic reconnection is ubiquitous in the solar wind across heliocentric distances and may provide new insights into the evolution of solar wind CS and the role of reconnection in solar wind turbulence.

**Acknowledgments**

All data used in this study are publicly available from the NASA Planetary Data System, Planetary Plasma Interactions Node. The MAVEN MAG calibrated data are available from the MAVEN MAG Calibrated Data Bundle (Connerney 2026), the MAVEN SWIA calibrated data from the MAVEN SWIA Calibrated Data Bundle (Halekas 2026), and the MAVEN SWEA calibrated data from the MAVEN SWEA Calibrated Data Bundle (Mitchell 2026). The Tianwen-1 MOMAG data sets are publicly available at https://moon.bao.ac.cn/web/zhmanager/mars1 and http://space.ustc.edu.cn/dreams/tw1_momag/. This work was partially supported by NASA grant NNH10CC04C through the MAVEN Project, NASA grants 80NSSC23K0911 and 80NSSC24K1843, DOE grant DE-SC0024639, the Alfred P. Sloan Research Fellowship, and the IBM Einstein Fellow Fund at the Institute for Advanced Study, Princeton. Part of this work is supported by the French space agency CNES for the observations obtained with the SWEA instrument.

Software: Data analysis were performed using PySPEDAS (Angelopoulos et al. 2026).

**Appendix A**

Figure A1 illustrates the ion velocity distribution functions (VDFs) along $V_L$ and in the $V_L$–$V_N$ plane for both events, comparing the ion VDFs within the exhaust regions with those in the surrounding regions. It can be seen that the ions undergo a clear bulk deceleration along $V_L$ within the exhaust regions, consistent with reconnection-driven plasma outflows.

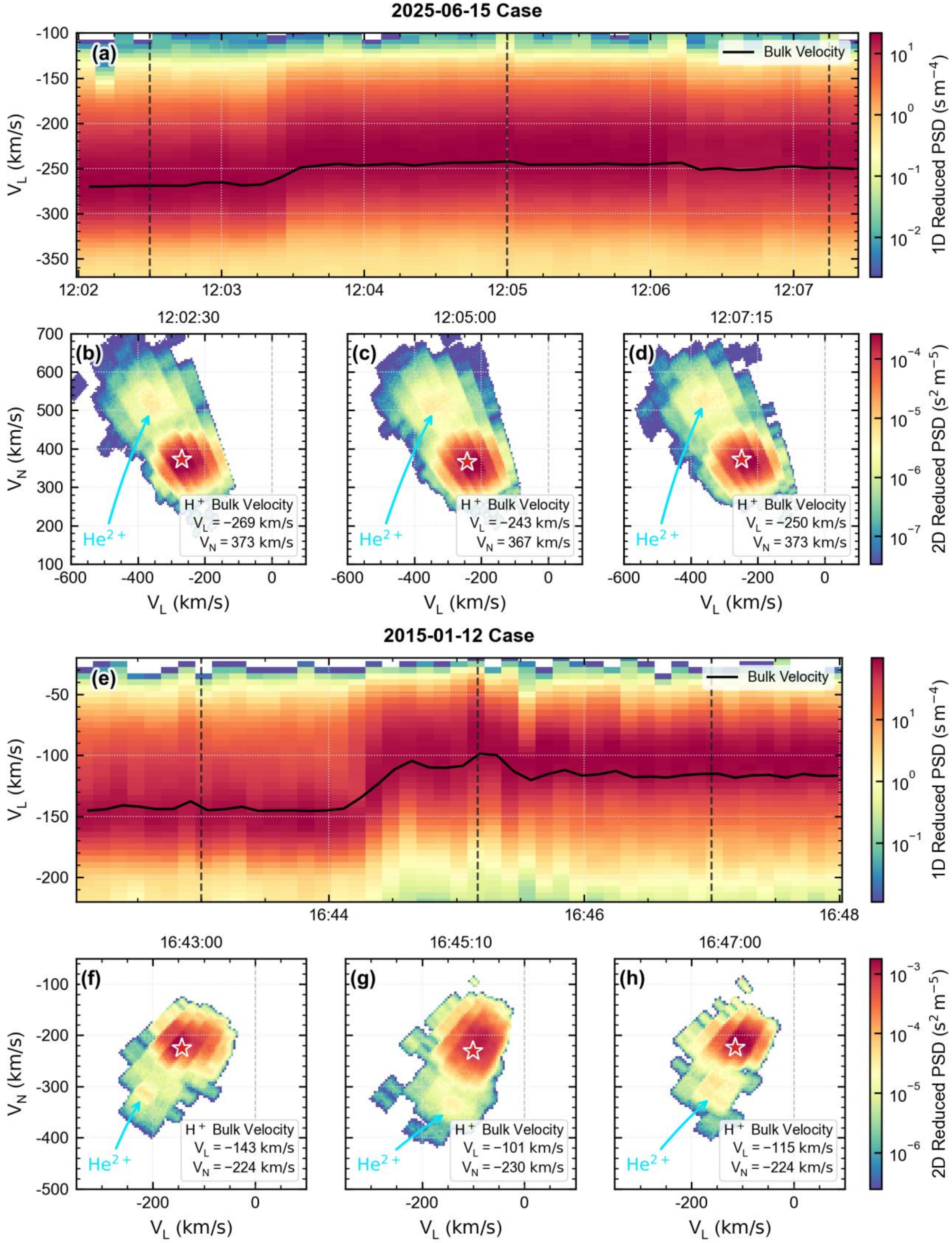
2025-06-15 Case
(a)
Bulk Velocity
$V_L$ (km/s)
1D Reduced PSD (s m$^{-4}$)
12:02
12:03
12:04
12:05
12:06
12:07
12:02:30
12:05:00
12:07:15
(b)
(c)
(d)
$V_N$ (km/s)
$V_L$ (km/s)
He$^{2+}$
H$^+$ Bulk Velocity
$V_L$ = −269 km/s
$V_N$ = 373 km/s
H$^+$ Bulk Velocity
$V_L$ = −243 km/s
$V_N$ = 367 km/s
H$^+$ Bulk Velocity
$V_L$ = −250 km/s
$V_N$ = 373 km/s
2D Reduced PSD (s$^2$ m$^{-5}$)
2015-01-12 Case
(e)
Bulk Velocity
16:44
16:46
16:48
16:43:00
16:45:10
16:47:00
(f)
(g)
(h)
H$^+$ Bulk Velocity
$V_L$ = −143 km/s
$V_N$ = −224 km/s
H$^+$ Bulk Velocity
$V_L$ = −101 km/s
$V_N$ = −230 km/s
H$^+$ Bulk Velocity
$V_L$ = −115 km/s
$V_N$ = −224 km/s

**Figure A1.** Ion velocity distribution functions for the 15 June 2025 and 12 January 2015 events. Panels (a) and (e) show the one-dimensional reduced ion phase space density (PSD) as a function of the $V_L$. The black solid lines indicate the calculated ion bulk velocity projected onto the $L$ direction. The vertical dashed lines mark the representative times selected for the two-dimensional distributions. Panels (b)–(d) and (f)–(h) present the corresponding two-dimensional reduced ion velocity distribution functions in the $V_L$–$V_N$ plane at the indicated times. The white stars denote the $H^+$bulk velocity. The cyan arrows indicate the $He^{2+}$.

## Appendix B

Figure B1 illustrates the electron energy pitch-angle distributions for both events. For both cases, the 30–300 eV electrons exhibit clear field-aligned distributions outside the exhaust regions, whereas the distributions become more isotropic within the exhaust regions.

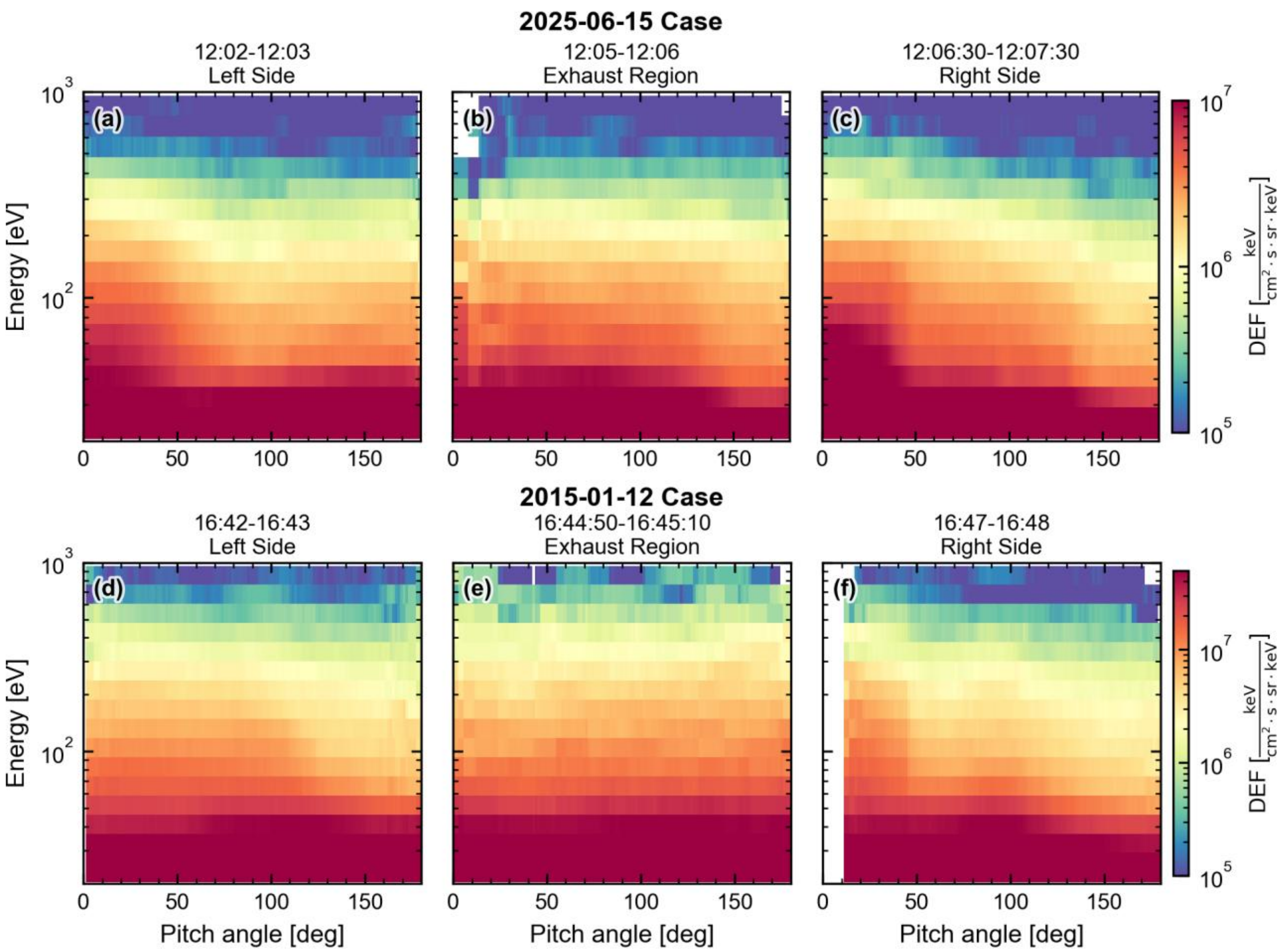

**Figure B1.** Electron energy–pitch angle distributions for the 15 June 2025 and 12 January 2015 reconnection events. Panels (a)–(c) show the electron differential energy flux (DEF) for the 15 June 2025 event during the left side of the current sheet, within the reconnection exhaust region, and on the right side of the current sheet, respectively. Panels (d)–(f) show the corresponding distributions for the 12 January 2015 event.

## Appendix C

Figure C1 shows simultaneous observations of the first CS event by MAVEN and Tianwen-1 on 15 June 2025. The two spacecraft, separated by approximately $4.5\,R_M$, observed similar magnetic field signatures associated with the CS. These observations indicate that the CS extended over at least several Martian radii, consistent with the large spatial scale inferred from the crossing duration. Given that this scale exceeds the characteristic size of the Martian induced magnetosphere (several $R_M$), the results suggest that such structure can influence the Martian induced magnetosphere on a global scale.

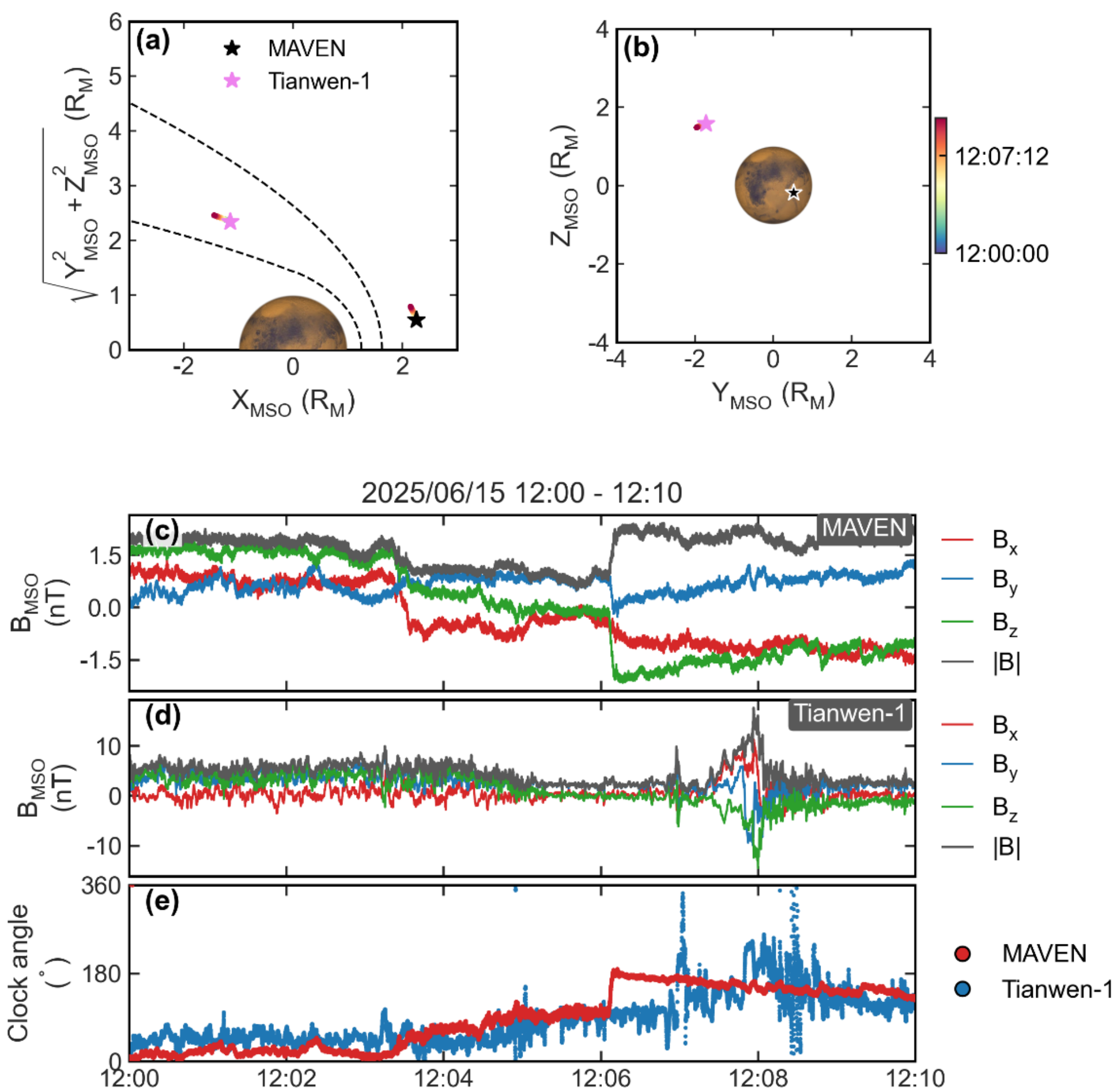

**Figure C1.** Simultaneous observations of the CS by MAVEN and Tianwen-1 on 15 June 2025. Panels (a) and (b) show the spacecraft locations in the $X_{\mathrm{MSO}} - \sqrt{Y_{\mathrm{MSO}}^2 + Z_{\mathrm{MSO}}^2}$ and $YZ_{\mathrm{MSO}}$ planes. The black and magenta stars denote the positions of MAVEN and Tianwen-1, respectively. Panels (c) and (d) present the magnetic field components measured by MAVEN and Tianwen-1 from 12:00 to 12:10 UT. (e) shows the clock angles derived from the magnetic field measurements.